\documentclass[twocolumn]{aastex631}

\newcommand{\angstrom}{\textup{\AA}}

\newcolumntype{b}{X}
\newcolumntype{s}{>{\hsize=.5\hsize}X}

\usepackage{textcomp}
\usepackage{CJKutf8}
\usepackage{amsmath}
\usepackage{color,soul}
\usepackage{tabularx}

\begin{document}

\title{An Ultramassive White Dwarf with a Likely Oxygen-Neon Core}

\correspondingauthor{Stefan M. Arseneau}
\author[0000-0002-6270-8624]{Stefan M. Arseneau}
\email{arseneau@bu.edu}
\affiliation{Department of Astronomy \& Institute for Astrophysical Research, Boston University, 725 Commonwealth Ave., Boston, MA 02215, USA}

\author[0000-0001-5941-2286]{J.~J.~Hermes}
\affiliation{Department of Astronomy \& Institute for Astrophysical Research, Boston University, 725 Commonwealth Ave., Boston, MA 02215, USA}

\author[0000-0002-0572-8012]{Vedant Chandra}
\affiliation{Center for Astrophysics $\mid$ Harvard \& Smithsonian, 60 Garden St, Cambridge, MA 02138, USA}

\author[0000-0002-9090-9191]{Roberto Raddi}
\affiliation{Departament de F\'{i}sica, Universitat Polit\`{e}cnica de Catalunya, c/Esteve Terrades 5, 08860 Castelldefels, Spain}

\author[0000-0002-3524-190X]{Maria E. Camisassa}
\affiliation{Departament de F\'{i}sica, Universitat Polit\`{e}cnica de Catalunya, c/Esteve Terrades 5, 08860 Castelldefels, Spain}

\author[0000-0002-6153-7173]{Alberto Rebassa-Mansergas}
\affiliation{Departament de F\'{i}sica, Universitat Polit\`{e}cnica de Catalunya, c/Esteve Terrades 5, 08860 Castelldefels, Spain}

\author[0000-0001-5777-5251]{Santiago Torres}
\affiliation{Departament de F\'{i}sica, Universitat Polit\`{e}cnica de Catalunya, c/Esteve Terrades 5, 08860 Castelldefels, Spain}

\begin{abstract}

The core composition of ultramassive white dwarfs remains an open question in stellar evolution. The carbon content of white dwarf cores is critical to their role as progenitors of Type Ia supernovae. However, because the stellar photosphere only extends to the outermost layer of the star, observational probes of core compositions are limited. Here we present gravitational redshift measurements of an ultramassive white dwarf, SDSS J060851.44-005950.3, which indicate the likely presence of an oxygen-neon core. We measure the mass ($1.226_{-0.025}^{+0.024} M_\odot$) and radius ($0.491_{-0.009}^{+0.009}~R_\oplus$) of the white dwarf using gravitational redshifts from high-resolution UVES and MagE spectra paired with independent constraints from photometry. By comparing to state-of-the-art mass-radius relations for ultramassive white dwarfs, we find preference for a oxygen-neon core over a carbon-oxygen core, with a Bayes factor of $2.7$. This is a white dwarf which is likely structurally incapable of producing a Type Ia supernova, according to current understanding of supernova physics. This object provides evidence that white dwarfs which pass through the Q-branch without experiencing a delay in cooling compared to the normal white dwarf cooling sequence likely have oxygen-neon cores.

\end{abstract}

\section{Introduction} \label{sec:intro}

Ultramassive white dwarfs with masses $>1.2$~$M_\odot$ are predicted to form from single progenitors with zero-age main sequence masses of $7-9.5$~$M_\odot$ \citep{2024ApJS..270...29L}, for which core temperatures during the asymptotic giant branch phase of stellar evolution are high enough to initiate carbon burning. This results in white dwarfs with degenerate cores which consist primarily of $^{16}\text{O}$ and $^{20}\text{Ne}$ (O/Ne; \citealt{2007A&A...476..893S, 2018MNRAS.480.1547L}). Most white dwarfs with masses above $\approx 1.05$~$M_\odot$ which form via single star evolution are predicted to have O/Ne cores \citep{2015MNRAS.446.2599D}, although the exact boundary remains unclear. (\citealt{2021A&A...646A..30A} have found that it may still be possible to form C/O cores via single-star evolution; see also \citealt{1996ApJ...472..783D}.) 

Further complicating this picture, about 50\% of white dwarfs with masses $>0.90~M_\odot$ are expected to be merger remnants \citep{2012A&A...546A..70T, 2020ApJ...891..160C, 2023MNRAS.518.2341K, 2024ApJ...974...12J}, identified from among other things the presence of magnetism, rapid rotation periods, and high transverse velocities, combined with population synthesis modeling \citep{2020A&A...636A..31T}. These objects may have never reached sufficient temperatures to undergo carbon burning, and therefore could host C/O cores \citep{2014ARA&A..52..107M, 2022MNRAS.512.2972W}. However, \cite{2021ApJ...906...53S} have found that this may not be the case, and indicate that merger products may produce O/Ne cores when the final mass is $\gtrsim1.05~M_\odot$. 

\cite{2019Natur.565..202T} observed an over-density of ultramassive white dwarfs on the Gaia color-magnitude diagram, known as the Q-branch. Some objects on the Q-branch ($\approx 8\%$) exhibit abnormally large transverse velocities, many of which would require an $8$~Gyr delay in cooling to explain in excess of the delays expected of crystallization or binary merger delay time \citep{2019ApJ...886..100C}. A leading hypothesis to explain their over-density is due to buoyant $^{22}$Ne crystals distilling in an inner C/O core and floating to the outer core, transferring gravitational potential energy into heat, delaying cooling \citep{2024Natur.627..286B}. In white dwarfs with O/Ne cores, $^{22}$Ne couples with $^{20}$Ne during crystallization, and so the crystallization proceeds normally from the inner core out with no significant cooling delay relative to the normal white dwarf cooling sequence \citep{2025ApJ...991...64C}. The presence of these white dwarfs on the Q-branch with anomalously high kinematics are thus indirect observational signatures of the presence of ultramassive merger products with C/O cores (see also \citealt{2021ApJ...911L...5B,2023ApJ...955L..33S}). 

\cite{2020NatAs...4..663H} detected atomic carbon absorption lines in the spectrum of an ultramassive ($1.140\pm0.008~M_\odot$) white dwarf with high velocity relative to the local standard of rest. In addition, their spectroscopic analysis displays no trace of helium in the spectrum of the star. They concluded that this was a Q-branch merger remnant, and that the atomic carbon lines were likely see-through core material detected through a particularly thin atmosphere of hydrogen and helium which was burnt away during the merger process. Still, because the high surface gravity of white dwarfs induces rapid sedimentation, observations of carbon in the spectrum are not sufficient to constitute direct observational evidence of a C/O core in the white dwarf.

O/Ne cores have been inferred in massive white dwarfs undergoing recurrent novae. These are cataclysmic variables, constituting a close binary system with a non-degenerate companion which fills its Roche lobe and undergoes stable mass transfer onto the white dwarf \citep{2008clno.book...77S}. Eventually, mass accretion onto the white dwarf can trigger a thermonuclear runaway of hydrogen burning on the surface of the white dwarf, ejecting the accreted material \citep{2012BASI...40..419S}. This can permit deeper regions of the star to be probed directly via spectroscopy. Recently, \cite{2026MNRAS.547ag361R} have identified Ne III and Ne V absorption lines in the spectrum of one such nova, indicating a white dwarf with an O/Ne core composition. However, the thermonuclear reactions associated with novae make inferring core composition challenging.

Constraining the core composition of massive white dwarfs has important implications for the search for Type Ia supernova progenitors. These supernovae are widely understood to be formed from runaway thermonuclear carbon detonation in the degenerate core of a white dwarf, meaning that only white dwarfs with cores rich in $^{12}\text{C}$ can form Type Ia supernovae \citep{2015ApJ...805L...6S}. When such conditions as would otherwise trigger a Type Ia supernova are achieved, white dwarfs with O/Ne cores are predicted to directly collapse into neutron stars via electron capture rather than form Type Ia supernovae \citep{1991ApJ...367L..19N}. Massive white dwarfs with C/O cores represent the population of stars which can most easily be driven to the conditions of carbon ignition, and therefore are the most promising candidates in searches for the progenitors of Type Ia supernovae.

The most promise for direct measurement of core compositions in ultramassive white dwarfs has long come from asteroseismology. Several massive and hydrogen rich (spectral type DA) white dwarfs exhibit $g$-mode pulsations \citep{2005A&A...432..219K, 2010MNRAS.405.2561C, 2013MNRAS.430...50C, 2013ApJ...771L...2H, 2017MNRAS.468..239C, 2019MNRAS.486.4574R}. Asteroseismic models can in principle distinguish between a C/O and O/Ne core composition \citep{2019AA...621A.100D,2021A&A...646A..30A}. In practice though, the limited number of pulsation modes (typically 3-5 modes per star, such as in \citealt{2009MNRAS.396.1709C}) detectable in most variable white dwarfs makes this measurement challenging. Furthermore, ultramassive white dwarfs are significantly crystallized by the time they enter the DAV instability strip \citep{2013ApJ...779...58R, 2019AA...632A.119C,2025ApJ...994..255J}. Poor mode penetration into the crystallized core means that constraining interior physics with seismology is difficult.

Recently, \cite{2025ApJ...980L...9D} analyzed 19 pulsation modes of WD J$0135+5722$, finding a mass of either $1.118\pm0.002~M_\odot$ in the case that the star has a O/Ne core or $1.135\pm0.004~M_\odot$ in the case of a C/O core. Likewise, \cite{2025ApJ...988...32C} analyzed 13 pulsation modes of WD J$0049-2525$, finding a mass of $\geq 1.29$, assuming an O/Ne core. %Additionally, \cite{2025ApJ...994..255J} have performed photometric followup of 31 pulsating white dwarfs with $M>0.9M_\odot$, but did not attempt to measure the core composition.

Due to their high densities, white dwarfs exhibit typical gravitational redshifts of $\approx 32$~km s$^{-1}$ (for a $0.6~M_\odot$ white dwarf; \citealt{2010ApJ...712..585F, 2020ApJ...899..146C}). Gravitational redshift is a relativistic effect in which photons leaving a star's gravitational potential well undergo a redshift of
\begin{equation}
    v_\text{g} = \frac{GM}{cR}
\end{equation}
where $M$ is the stellar mass, $c$ is the speed of light, and $R$ is the stellar radius. For individual white dwarfs, gravitational redshift is perfectly degenerate with the true radial velocity of the star. When it can be measured, it provides a valuable constraint on the mass-radius relation, which is largely independent of model spectra.

Wide binary stars represent a reliable method for measuring the gravitational redshift of individual white dwarfs. By comparing the apparent radial velocity of white dwarfs in these systems to that of their main sequence companion, gravitational redshift can be isolated for a single star, providing a constraint on the mass-radius relation \citep{2024ApJ...963...17A, 2025A&A...695A.131R}. 

The third data release of Gaia has allowed the identification of over 22,000 candidate wide binary systems containing one white dwarf and one main sequence star \citep{2021MNRAS.506.2269E, 2026arXiv260415939R}. Using the main sequence star's radial velocity can then provide accurate constraints on the mass-radius relation of white dwarfs, which can be used to probe stellar structure. For white dwarfs of typical mass ($\approx 0.6~M_\sun$), the mass-radius relation is sensitive to the temperature of the star and the mass of the hydrogen envelope \citep{2024ApJ...977..237C, 2026ApJ..1000..297A}. For the most massive white dwarfs, though, the mass-radius relation is most sensitive to the composition of the core, and can therefore distinguish between C/O and O/Ne cores at high precision \citep{2019A&A...625A..87C, 2022MNRAS.511.5198C}. 

In this work, we present a measurement of the core composition of SDSS J060851.44-005950.3 (hereafter referred to as SDSS J0608$-$0059), an ultramassive white dwarf with a common-proper-motion main sequence companion for which gravitational redshift can be measured from high-resolution spectroscopy. The projected separation of the binary system is $2684$~au, sufficiently wide to ensure they have never interacted in the past. We describe our spectroscopic observations and our methods for measuring the white dwarf mass and radius in Section \ref{sec:methods}, and in Section \ref{sec:results} we present our findings.

\section{Methods} \label{sec:methods}

\begin{figure}
    \centering
    \includegraphics[width=\linewidth]{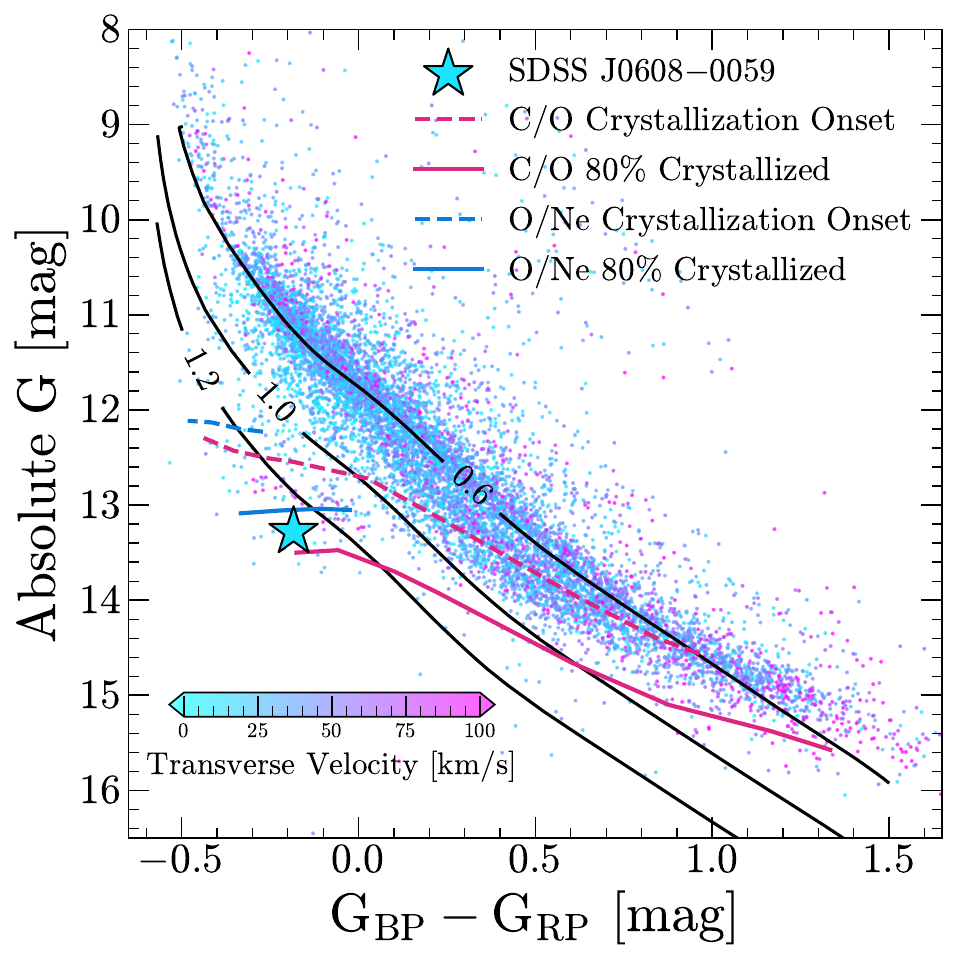}
    \caption{SDSS J0608$-$0059 sits well below the main white dwarf cooling track, indicating that it is ultramassive. It has low kinematics, with a transverse velocity of $10.84\pm0.06$~km s$^{-1}$ according to Gaia astrometry. Evolutionary tracks for white dwarfs with masses of $0.6$~$M_\odot$, $1.0$~$M_\odot$, and $1.2$~$M_\odot$ are marked in black (using C/O core evolutionary tracks from \citealt{2020ApJ...901...93B}), and stars are colored by their transverse velocities calculated from Gaia astrometry. Plotted in purple are boundaries corresponding to the onset of crystalllization (dashed) and 80\% crystallization (solid) for white dwarfs with C/O cores. The same boundaries for white dwarfs with O/Ne cores are marked in blue \citep{2024A&A...683A.101C}. Its position on the color magnitude diagram indicates that SDSS J0608$-$0059 is mostly crystallized.}
    \label{fig:cmd}
\end{figure}

SDSS J0608$-$0059 has an apparent magnitude of $17.2$ in the Gaia G-band. Figure \ref{fig:cmd} displays the star's position in the color-magnitude diagram (CMD), along with crystallization boundaries for C/O and O/Ne core compositions from \cite{2024A&A...683A.101C}, indicating that SDSS J0608$-$0059 is likely mostly crystallized, and has evolved through the Q-branch overdensity where some white dwarfs have very long cooling delays \citep{2019ApJ...886..100C}. Such objects are rare in the solar neighborhood (SDSS J0608$-$0059 is within 100 pc), as white dwarfs cool rapidly following crystallization.

\subsection{Gravitational Redshift}

\begin{figure*}[t]
    \centering
    \includegraphics[width=0.47\linewidth]{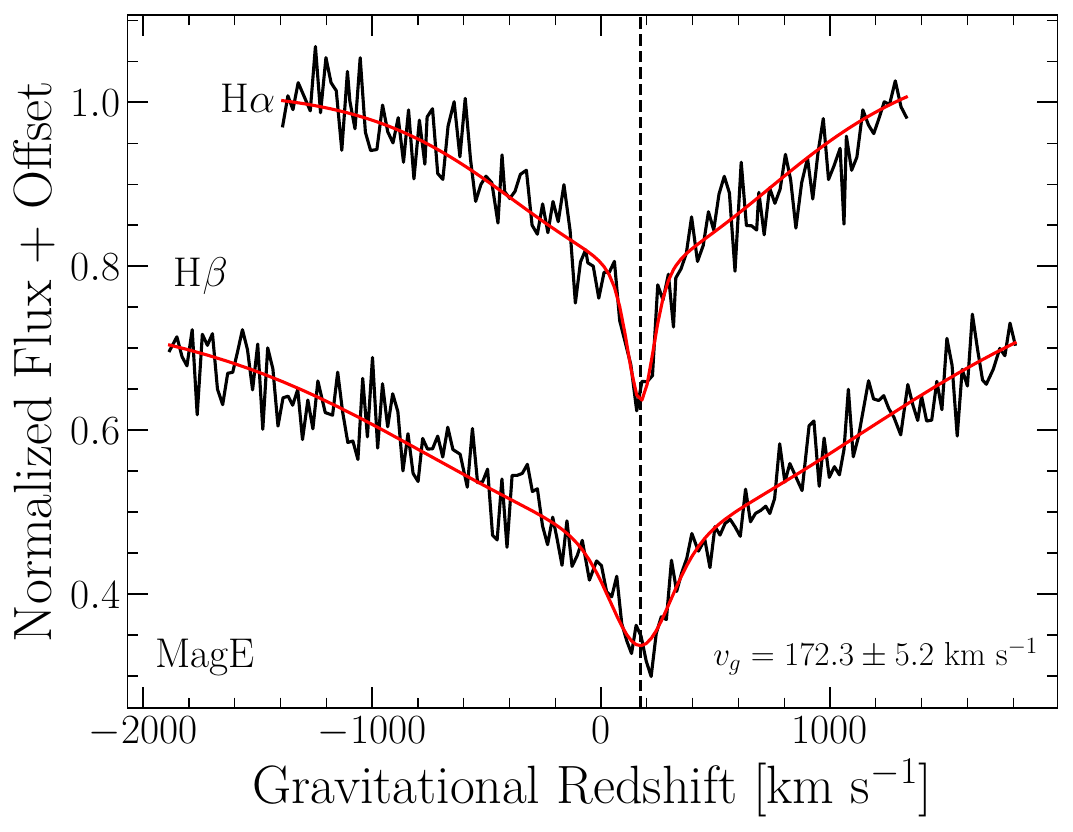}
    \includegraphics[width=0.47\linewidth]{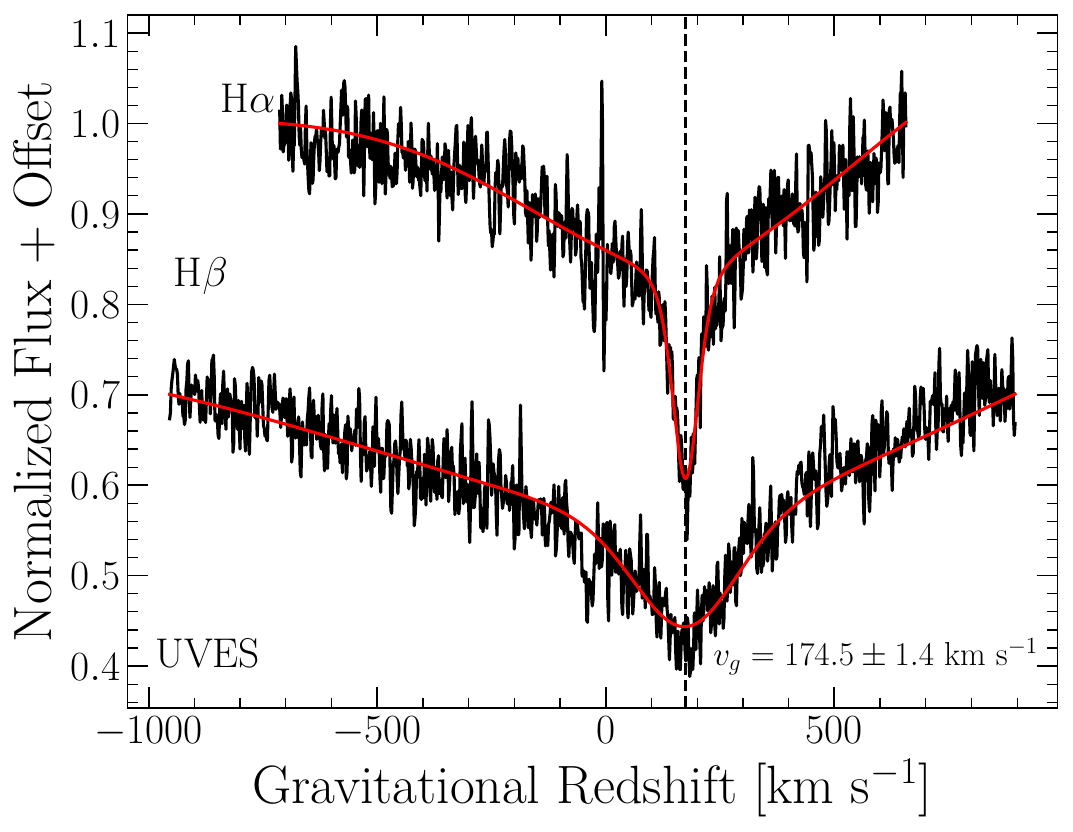}
    \caption{Best fit gravitational redshift to the H$\alpha$ and H$\beta$ lines of SDSS J0608$-$0059 in a window of $\pm30$~$\angstrom$ for the lower-resolution MagE data (left) and $\pm15$~$\angstrom$ for the higher-resolution UVES data (right). The gravitational redshift of the white dwarf is its radial velocity in the frame of reference of the common proper motion companion, corrected for the orbital motion and gravitational redshift of the companion. Therefore the measured radial velocity is due to the gravitational redshift of SDSS J0608$-$0059. The centroid radial velocity is marked as a horizontal line.}
    \label{fig:lines}
\end{figure*}

On 2024 April 01 we observed SDSS J0608$-$0059 using the MagE spectrograph \citep{2008SPIE.7014E..54M} on the $6.5$~m Magellan Clay Telescope at Las Campanas Observatory, Chile for one exposure at a $0.7$~arcsecond slit for an exposure time of $1200$~s at a spectral resolution of $R=4100$ ($1.6$~$\angstrom$ at H$\alpha$). We reduced the spectrum using version \texttt{1.15.0} of the python package \texttt{PypeIt}. Our observed spectrum has a signal to noise ratio per pixel of $53$. 

Additionally, we observed SDSS J0608$-$0059 twice with the Ultraviolet and Visual Eschelle Spectrograph (UVES) on the European Southern Observatory's $8.2$~m Very Large Telescope (ESO VLT; \citealt{2000SPIE.4008..534D}). The white dwarf was originally observed on 2022 January 12 for a single $3600$~s exposure, at a signal to noise ratio per pixel of $5$. Two additional observations were performed, for $3980$~s on 2025 September 12 and $3600$~s on 2025 September 13, at signal to noise ratios per pixel of $29$ and $34$. All observations were made with a $2.2$~arcsecond slit, yielding a spectral resolution of $R = 18,500$~$\angstrom$ ($0.35$~$\angstrom$ at H$\alpha$), and were reduced with the standard ESO pipeline. We first use \texttt{astropy} \citep{2013A&A...558A..33A, 2018AJ....156..123A, 2022ApJ...935..167A} to apply a Doppler shift to each observation to bring it into the heliocentric reference frame, and then correct the spectrum to vacuum wavelengths via the transformation of \cite{1991ApJS...77..119M}. Finally, we coadd each of the three UVES observations. The coadded spectrum has an average signal to noise ratio per pixel of $46$. 

We perform a spectroscopic fit to the H$\alpha-\zeta$ lines of the MagE spectrum using the 1D NLTE models of \cite{2006ApJ...651L.137K, 2009ApJ...696.1755T, 2011ApJ...730..128T}. The best-fitting atmospheric parameters are $T_\text{eff} = 17,110\pm210$~K and $\log g = 9.03\pm0.02$~dex. The same fit to our UVES spectrum yields $T_\text{eff} = 20,000\pm 240$~K and $\log g = 8.94\pm0.02$~dex. The difference between these two measurements is likely related to unreliable flux calibration in the spectrum. In our subsequent analysis, we do not rely on spectroscopically determined atmospheric parameters.

We measure the gravitational redshift of SDSS J0608$-$0059 by comparing the radial velocity measured from spectroscopy to the radial velocity of its main sequence companion. We measure the radial velocity of SDSS J0608$-$0059 from UVES and MagE spectra separately using the python package \texttt{corv}\footnote{\url{https://github.com/vedantchandra/corv}} \citep{2024ApJ...963...17A}, fitting two Voigt profiles to each of the H$\alpha$ and H$\beta$ lines simultaneously, with a fixed radial velocity applied to each line. For UVES we fit in a window of $\pm 15$~$\angstrom$ about the line centroid. This window is chosen to mitigate redshifts due to unmodeled physics in the deep stellar photosphere (and therefore further in the wings of the absorption lines; see \citealt{2020A&A...638A.131N, 2025ApJ...991..190A}). MagE spectroscopy only partially resolves the core of the hydrogen absorption lines, and so in order to maximize signal we fit in a window of $\pm 30$~\angstrom. Two of the UVES exposures as well as the MagE spectrum have high enough signal to noise that we are able to measure their individual radial velocities. From these, we determine radial velocities of $203.5\pm 1.3$~km s$^{-1}$, $202.7 \pm 1.4$~km s$^{-1}$, and $201.1 ± 3.3$~km s$^{-1}$. Thus SDSS J0608$-$0059 does not appear radial-velocity variable, and likely is not an unresolved close binary system. 

The radial velocity of the main sequence companion measured by the Gaia radial velocity spectrometer \citep{2023A&A...674A...5K} is $28.80\pm0.78$~km s$^{-1}$. The temperature and surface gravity of the main sequence companion are reported in the Gaia archive as $T_\text{eff} = 3963$ and $\log g = 5.03$~dex. We determine a mass and radius from these parameters using MIST isochrones, assuming solar metallicity \citep{2016ApJS..222....8D, 2016ApJ...823..102C, 2026ApJS..283...64D, 2026ApJS..283...41B}. From this we calculate the gravitational redshift of the main sequence companion as $0.72$~km s$^{-1}$, and correct the radial velocity of the main sequence star. The corrected radial velocity is $28.1\pm0.8$~km s$^{-1}$. 

Assuming that the mass of the white dwarf is $1.26$~$M_\odot$ based on its position in the color-magnitude diagram (\citealt{2021MNRAS.508.3877G} report $1.26\pm0.01~M_\odot$ from a fit to Gaia photometry, a value we will refine later), we estimate that the system has a Keplerian orbital velocity of $0.67$~km s$^{-1}$. The Gaia-measured transverse velocity of the main sequence companion relative to SDSS J0608$-$0059 is $0.22$~km s$^{-1}$, meaning that the radial velocity component of the differential orbital motion is $\pm 0.63$~km s$^{-1}$. Because the direction of motion cannot be determined from Gaia astrometry alone, we add this value in quadrature to the uncertainty of the corrected radial velocity. This results in a corrected main sequence radial velocity of $28.1\pm 1.0$~km s$^{-1}$. 

We determine the gravitational redshift of the white dwarf by subtracting this velocity from its best-fit radial velocity, and we combine their uncertainties in quadrature. From the UVES spectrum we measure a gravitational redshift of $174.5 \pm 1.4$~km s$^{-1}$, and from MagE we measure $173.0 \pm 3.5$~km s$^{-1}$. Figure \ref{fig:lines} presents the fits to each sets of lines. Our adopted gravitational redshift is the inverse variance weighted average of these two measurements, $174.5 \pm 1.3$~km s$^{-1}$.

\subsection{The Spectral Energy Distribution}

\begin{figure}[t]
    \centering
    \includegraphics[width=\linewidth]{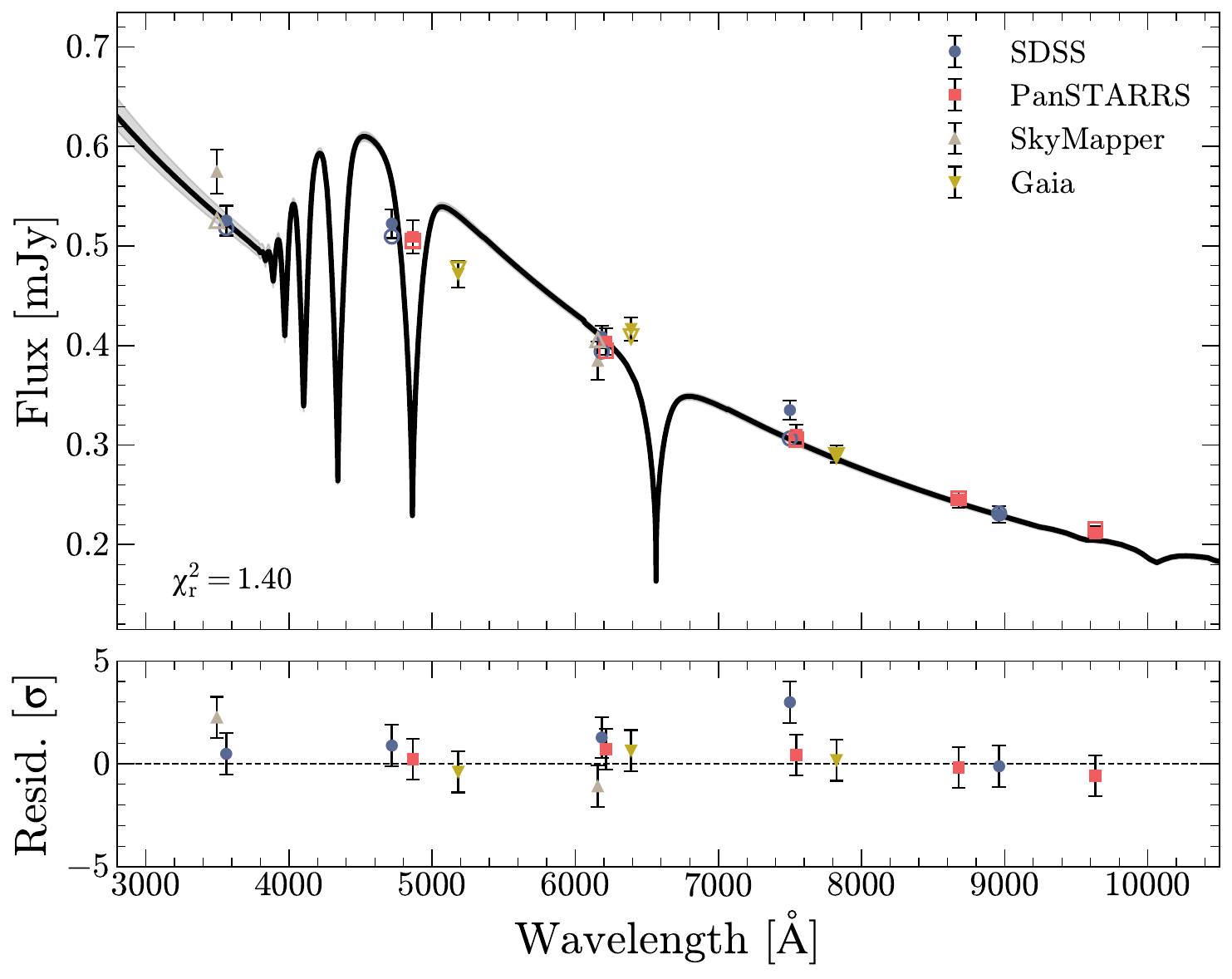}
    \caption{Model atmosphere best fits to Gaia, SDSS, PanSTARRs, and SkyMapper photometry. We find a best fit temperature of $17,790_{-370}^{+400}$~K. The best fit model atmosphere is plotted in black, with photometry computed from the model spectrum in each band plotted as open points. We find a reduced chi-square statistic ($\chi^2_r$) of $1.4$. The $16^\text{th}$ to $84^\text{th}$ confidence interval sampled from the MCMC posteriors is shaded in gray; the ultraviolet region is the most poorly constrained. The bottom panel presents residuals between the observed photometry and photometry computed from the spectrum corresponding to the median of the MCMC posterior distribution in units of standard deviations, showing no structure.}
    \label{fig:bestfit}
\end{figure}

With gravitational redshift measured, one additional measurement is necessary to fully constrain the mass-radius relation of SDSS J0608$-$0059. We compute posterior distributions for mass, radius, effective temperature, distance, and extinction by combining our measured gravitational redshift with photometry from Gaia \citep{2016A&A...595A...1G, 2018A&A...616A...1G, 2021A&A...649A...1G, 2023A&A...674A..34G}, SDSS \citep{2007ApJS..172..634A, 2026AJ....171...52K}, PanSTARRs \citep{2016arXiv161205560C}, and SkyMapper \citep{2024PASA...41...61O}. We apply an offset of $-0.04$ and $+0.02$ to the SDSS $u$ and $z$ bands to put them onto the AB magnitude system \citep{2006AJ....132..676E}. To ensure that our photometry is reliable, we require SDSS bands not be flagged as any of \texttt{EDGE}, \texttt{PEAKCENTER}, \texttt{SATUR}, or \texttt{NOTCHECKED}, and we require that SkyMapper and PanSTARRs photometry have no flags set. \cite{2022ApJ...934..148H, 2024ApJ...969...68H} have found that this produces the most robust results. This removes only SkyMapper $v$-band photometry, leaving us with Gaia photometry, SDSS $ugriz$, PanSTARRs $grizy$, and SkyMapper $ugriz$. To account for systematic uncertainties in survey calibration and zero-points, we add $0.03$~mag of uncertainty as a noise floor in each band. 

We generate a synthetic spectral energy distribution by convolving 1D DA NLTE model spectra \citep{2006ApJ...651L.137K, 2009ApJ...696.1755T, 2011ApJ...730..128T} by the throughputs of each filter. These model spectra are valid in the range $6.5~\text{dex} < \log g < 9.5~\text{dex}$ and $1500\text{~K} < T_\text{eff} < 140,000\text{~K}$. This is converted to observed flux via the expression
\begin{equation}
    f_i = 4\pi H_i\left(T_\text{eff}, \log \frac{GM}{R^2}\right)\left(\frac{R}{d}\right)^2 10^{-0.4 \varepsilon_i A_V}
\end{equation}
where $H_i$ is the surface flux convolved from the synthetic spectrum in the $i$th filter band, $R$ is the radius, $M$ is the mass of the white dwarf, $d$ is distance to the star, $A_V$ is the $V$-band extinction, and $\varepsilon_i$ is a conversion factor determined using Table 6 of \cite{2011ApJ...737..103S}. 

\begin{figure*}[ht]
    \centering
    \includegraphics[width=\linewidth]{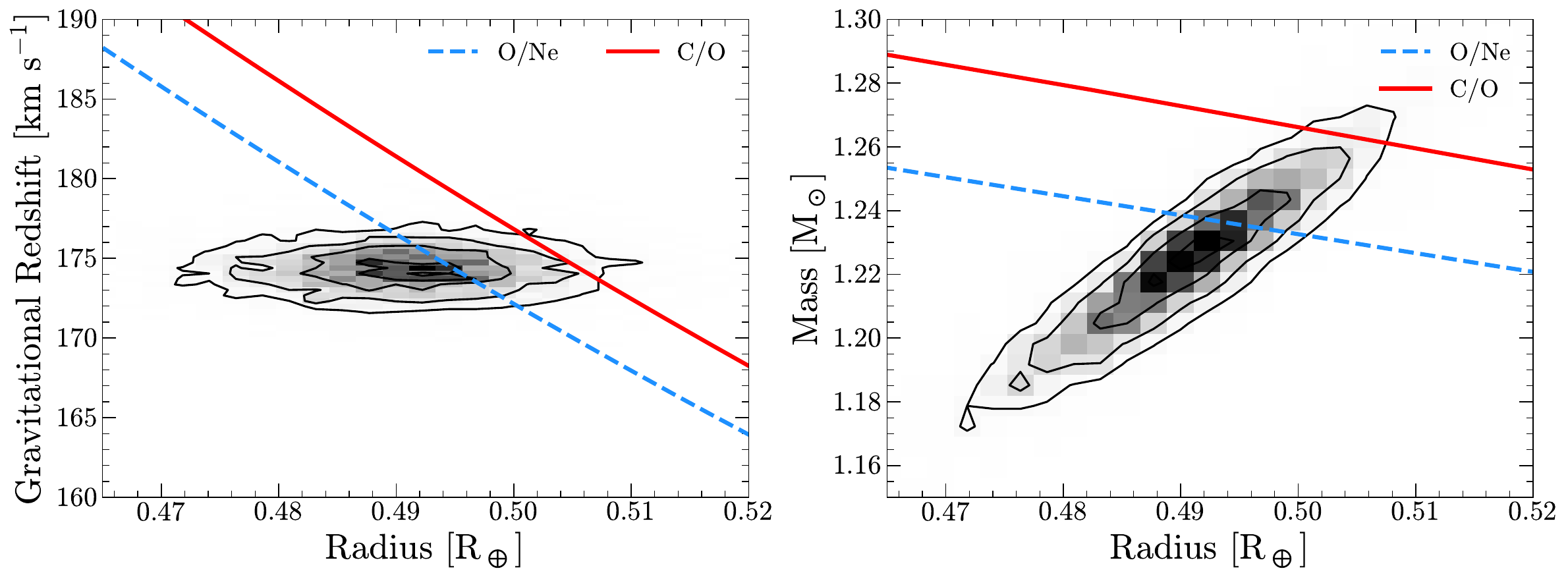}
    \caption{The posterior distributions of gravitational redshift and radius as well as mass and radius for SDSS J0608$-$0059. Contours represent the $0.5\sigma$, $1.0\sigma$, $1.5\sigma$, and $2.0\sigma$ confidence intervals for a 2D Gaussian. We observe that the oxygen/neon core models (blue) are preferred over the carbon-oxygen models (red) with a Bayes factor of $2.7$. We interpret this as evidence that ultramassive white dwarfs which pass through the Q-branch without experiencing a delay have O/Ne cores.}
    \label{fig:posterior}
\end{figure*}

We define a parameterized model with unknown parameters $\vec{\theta} = \{T_\text{eff}, R, d, A_v, M\}$ corresponding to effective temperature, stellar radius, distance, $V$-band extinction, and stellar mass. We calculate $\log g = \log \frac{GM}{R^2}$, applying Gaussian priors on $A_V$, $v_g$, and fluxes $\{f_i\}_{i=1}^n$, with a $5\%$ uncertainty for $A_V$. We apply a Gaussian prior on the parallax of the system, using the weighted mean of the Gaia-measured parallaxes of the white dwarf and its companion, with the addition of a $d^2$ term as outlined by \cite{2018AJ....156...58B, 2021AJ....161..147B}. We also require that $1500\text{~K} < T_\text{eff} < 140,000\text{~K}$ and $6.5~\text{dex} < \log g < 9.5~\text{dex}$ with uniform priors. The full likelihood function describing priors on all quantities, as well as the corresponding posterior distribution, are presented in Appendix \ref{app:a}. 

We calculate the posterior distribution via Markov Chain Monte Carlo (MCMC) with the python package \texttt{emcee} \citep{2013PASP..125..306F} using 50 walkers. We run a $5000$ step MCMC chain, which is enough to achieve convergence, and discard the first $1000$ steps of the resulting chain to remove burn-in. The best fit distribution of model spectra is presented in Figure \ref{fig:bestfit}. The model corresponding to the median of the posterior distribution fits the data with reduced chi-square of $1.4$. The posterior distribution is presented in the Appendix as Figure \ref{fig:corner}. A summary of our inferred parameters are presented in Table \ref{tab:exo}.

\subsection{Kinematic Analysis}

We assess the consistency of SDSS J0608$-$0059 as a product of single-star evolution by investigating its kinematics. Stellar mergers reset the cooling age of a white dwarf, and so white dwarfs formed via single-star evolution are intrinsically younger than those formed as merger products at the same mass and effective temperature. An intrinsically older population of white dwarfs has had more time to experience velocity increases due to gravitational scattering, especially as their orbits pass through the Galactic disc \citep{2009A&A...501..941H, 2012MNRAS.426..427W}. If SDSS J0608$-$0059 were a merger product, and hence intriniscally older than it appears, it would likely have abnormally large kinematics.

We adopt a true space radial velocity of $28.1\pm0.8$~km s$^{-1}$ (the kinematic radial velocity of the co-moving companion) and calculate three-dimensional velocities from Gaia proper motions and the median of the white dwarf distance posterior. We compare this to three-dimensional velocities from the catalog of \cite{2022A&A...658A..22R}, adopting the radial velocities reported in that catalog and calculating tangential velocities from Gaia proper motions with distances from \cite{2021AJ....161..147B}. Compared to the distribution of velocities from \cite{2022A&A...658A..22R}, SDSS J0608$-$0059 lies in the 35th percentile. Thus, there is no kinematic evidence that it is a merger byproduct with faster-than-expected kinematics. It does not appear to currently be or have ever been a delayed Q-branch white dwarf \citep{2026ApJ...999..146O}.

Additional circumstantial evidence for a non-merger origin come from the non-detection of a magnetic field and the existence of this object as a wide binary. \cite{2019ApJ...886..100C} and \cite{2025ApJ...991..226H} have found that the natal kick of order $\sim1$~km s$^{-1}$ associated with binary mergers is very often sufficient to disrupt weakly gravitationally bound wide binaries. While all available evidence is consistent with a non-merger origin, the evidence do not constitute definitive proof of a single-star origin; the possibility that this object is a merger product cannot be excluded.

\section{Discussion and Conclusions} \label{sec:results}

\begin{deluxetable}{llr}
  \label{tab:exo}
  \tablecaption{Measured parameters of SDSS\,J0608$-$0059. The likelihood function is provided in Appendix \ref{app:a}.}
  \tablehead{
    \colhead{} &
    \colhead{Parameter} &
    \colhead{Value}
  }
  \startdata
  Star Parameters   & $M$            & $1.226_{-0.025}^{+0.024}~M_\odot$ \\
                    & $R$            & $0.491_{-0.009}^{+0.009}~R_\oplus$ \\
                    & $T_\text{eff}$ & $17{,}820_{-370}^{+400}$~K \\
                    & $v_\text{g}$   & $174.5\pm 1.3$~km s$^{-1}$ \\
                    & $\log g$       & $9.225_{-0.009}^{+0.008}$~dex \\
  \tableline
  System Parameters & Distance          & $61.65_{-0.06}^{+0.06}$~pc \\
                    & $A_V$        & $(1\pm0.05)\times 10^{-3}$~mag \\
  \tableline
  Model Comparison  & $\mathcal{B} = \frac{p(\text{O/Ne})}{p(\text{C/O})}$     & $2.7$ \\
  \enddata
\end{deluxetable}
  
We plot the posterior distributions of gravitational redshift radius, and mass-radius relations in Figure \ref{fig:posterior}. In addition, we plot gravitational redshift-radius and mass-radius relations for ultramassive white dwarfs with carbon-oxygen and oxygen-neon core compositions from \cite{2022MNRAS.511.5198C} and \cite{2019A&A...625A..87C}, respectively. We find good agreement with models for a degenerate core consisting primarily of oxygen and neon. The posterior distribution prefers a core consisting of oxygen and neon with a Bayes factor of $2.7$, calculated as the ratio of the total posterior probability density integrated along the oxygen-neon and carbon-oxygen model curves. Similarly, we find that the posterior distribution indicates a $1.55\sigma$ rejection of a carbon-oxygen core. Importantly, because our parameters come from gravitational redshifts and measurements of the star's solid angle, our results are largely independent of the assumed mass-radius relation. 

%Corrected for its gravitational redshift, the star is in the 35th percentile of the three-dimensional white dwarf velocity distribution of \cite{2022A&A...658A..22R}. 

If it could be conclusively shown that SDSS J0608$-$0059 is not a merger remnant, our result would provide observational evidence for the conclusion that white dwarfs which are not formed from stellar mergers have O/Ne cores. As it stands, our result is consistent with the picture that white dwarfs that do not show extensive (Gyr) cooling delays in the Q-branch likely harbor O/Ne cores \citep{2024Natur.627..286B}.

Notably, the limiting factor in the statistical significance of our core composition measurement is our photometric radius constraints, not the gravitational redshift measurement. Combining the Gaia parallaxes of both of the binary companions allows us to reach a parallax uncertainty of $15~\mu$as (the apparent magnitude of the main-sequence companion in the G-band is $12.2$~mag), and improved astrometry from Gaia DR4 will likely bring our parallax uncertainty close to the $10~\mu$as noise floor. If additional UV data were collected, permitting a better constraint on the star's temperature, the core composition could likely be determined more conclusively.

%The signal of core composition differences in gravitational redshifts decreases with decreasing mass. At a mass of $1.226~M_\odot$, the maximum likelihood mass of this object, there is a difference in gravitational redshift of $3.05$~km s$^{-1}$ between the C/O and O/Ne models. Although the models of \cite{2019A&A...625A..87C, 2022MNRAS.511.5198C} do not extend to the $\approx1.05~M_\odot$ point where the transition from O/Ne to C/O core compositions are expected, 

Our findings are consistent with those previously reported for this system by \cite{2025A&A...695A.131R}, with improved uncertainties. Their reported gravitational redshift is $179.6\pm 7.0$~km s$^{-1}$, consistent with our finding of $174.5\pm 1.3$~km s$^{-1}$. They additionally report a mass and radius of $1.31\pm0.06~M_\odot$ and $0.51\pm0.1~R_\oplus$, which is comparable to our values of $1.226_{-0.025}^{+0.024}~M_\odot$ and $0.491_{-0.009}^{+0.009}~R_\oplus$. The measurement of \cite{2025A&A...695A.131R} was made using Gaia XP spectra, as well as UVES measurements at a signal-to-noise ratio per pixel of 5; hence our measurements, using survey photometry and a high signal-to-noise ratio spectrum, represent a sensible reduction in scatter relative to theirs. 

There are several effects which could in principle modify the mass-radius relation of this object in such a way as to affect our results, but none of them are a factor for this particular object. The thickness of the hydrogen envelope can induce a signal in the inferred gravitational redshift of up to $1$~km s$^{-1}$ in white dwarfs of $\approx 0.6~M_\odot$ \citep{2026ApJ..1000..297A}, but for a star as massive as SDSS J0608$-$0059, the mass-radius models of \cite{2019A&A...625A..87C, 2022MNRAS.511.5198C} predict an effect of $< 0.2$~km s$^{-1}$, which is not sufficient to substantially affect our results. Other potential modifications to the mass-radius relation --- including the exact value of the $^{12}$C$/^{16}$O or $^{16}$O$/^{20}$Ne ratios in the core, and the effects of general relativity \citep{2023MNRAS.523.4492A}, rotation, and magnetism --- would each produce changes to the measured gravitational redshift smaller than that induced by the hydrogen layer, and so are all negligible for this object. We find no spectroscopic or photometric evidence for the latter two. Our evidence for an O/Ne core composition is therefore insensitive to these systematic effects.

The generally accepted pathway for detonation of a Type Ia supernova is thermonuclear runaway C-burning in the core of a C/O white dwarf \citep{2012ApJ...748...35S}. For white dwarfs such as SDSS J0608$-$0059, conditions that in a C/O white dwarf might lead to the formation of a Type Ia supernova are instead likely to cause an accretion-induced collapse directly into a neutron star. This is because electron capture reactions with core Ne and Mg will sap the electron degeneracy pressure which supports the star, leading to gravitational collapse \citep{2010MNRAS.409..846D}. Stars with core compositions similar to SDSS J0608$-$0059 may be sites of $r$-process nucleosynthesis \citep{2014ApJ...794...28P, 2025ApJ...984..197B}; however, they are likely structurally incapable of producing Type Ia supernovae.

\section*{Acknowledgements}

We thank the anonymous referee for their constructive feedback, which has improved the quality of the manuscript. RR acknowledges support from Grant RYC2021-030837-I funded by MCIN/AEI/ 10.13039/501100011033 and by ``European Union NextGeneration EU/PRTR''. M.C. acknowledges grant RYC2021-032721-I, funded by MCIN/AEI/10.13039/501100011033 and by the European Union NextGenerationEU/PRTR. This research was partially supported by the Spanish MINECO grants PID2023-148661NB-I00 and by the AGAUR/Generalitat de Catalunya grant SGR-386/2021.

This work has made use of data from the European Space Agency (ESA) mission Gaia (\url{https://www.cosmos.esa.int/gaia}), processed by the Gaia Data Processing and Analysis Consortium (DPAC, \url{https://www.cosmos.esa.int/web/gaia/dpac/consortium}). Funding for the DPAC has been provided by national institutions, in particular the institutions participating in the \textit{Gaia} Multilateral Agreement.

This paper includes data gathered with the 6.5 meter Magellan Telescopes located at Las Campanas Observatory, Chile. Based on observations collected at the European Organisation for Astronomical Research in the Southern Hemisphere under ESO programs 115.28D4.001 and 0108.D-0328.

%\section*{Data availability}

%\vspace{5mm}





\bibliography{citations.bib}{}
\bibliographystyle{aasjournal}

\appendix
\restartappendixnumbering
\section{Corner Plot}  \label{app:a}

We use a likelihood function based on that of \cite{2025A&A...695A.131R} and \cite{2026ApJ..1000..297A}:
\begin{align}
    \ln \mathcal{L}(T_\text{eff}, R, d, A_V, M) &=  -\frac{1}{2}\sum_{i=1}^n\left[\frac{(f_{i,\text{obs}} - f_{i}(\theta))^2}{\sigma_{f_i}^2} + \ln 2\pi\sigma_{f_i}^2\right] - \frac{1}{2}\left[ \frac{(1/d - \varpi)^2}{\sigma_{\varpi}^2} + \ln 2\pi\sigma_{\varpi}^2\right] + 2\ln d \nonumber \\
    &-\frac{1}{2}\left[\frac{(A_V - A_{V,\text{NGF}})^2}{(0.05A_{V,\text{NGF}})^2} + \ln 2\pi (0.05A_{V,\text{NGF}})^2\right] \nonumber- \frac{1}{2}\left[\frac{(GM/Rc - v_\text{g})^2}{\sigma_{v_\text{g}}^2} + \ln 2\pi \sigma_{v_\text{g}}^2\right]
\end{align}
where $A_{V,\text{NGF}}$ is the $V$-band extinction reported by \cite{2021MNRAS.508.3877G} and $\varpi$ is the error-weighted average parallax of the white dwarf and its companion reported from Gaia astrometry. These correspond to Gaussian priors on all quantities except for distance, for which an additional geometric factor of $d^2$ is incorporated (see \citealt{2021AJ....161..147B}).

\begin{figure*}[h!]
    \centering
    \includegraphics[width=\linewidth]{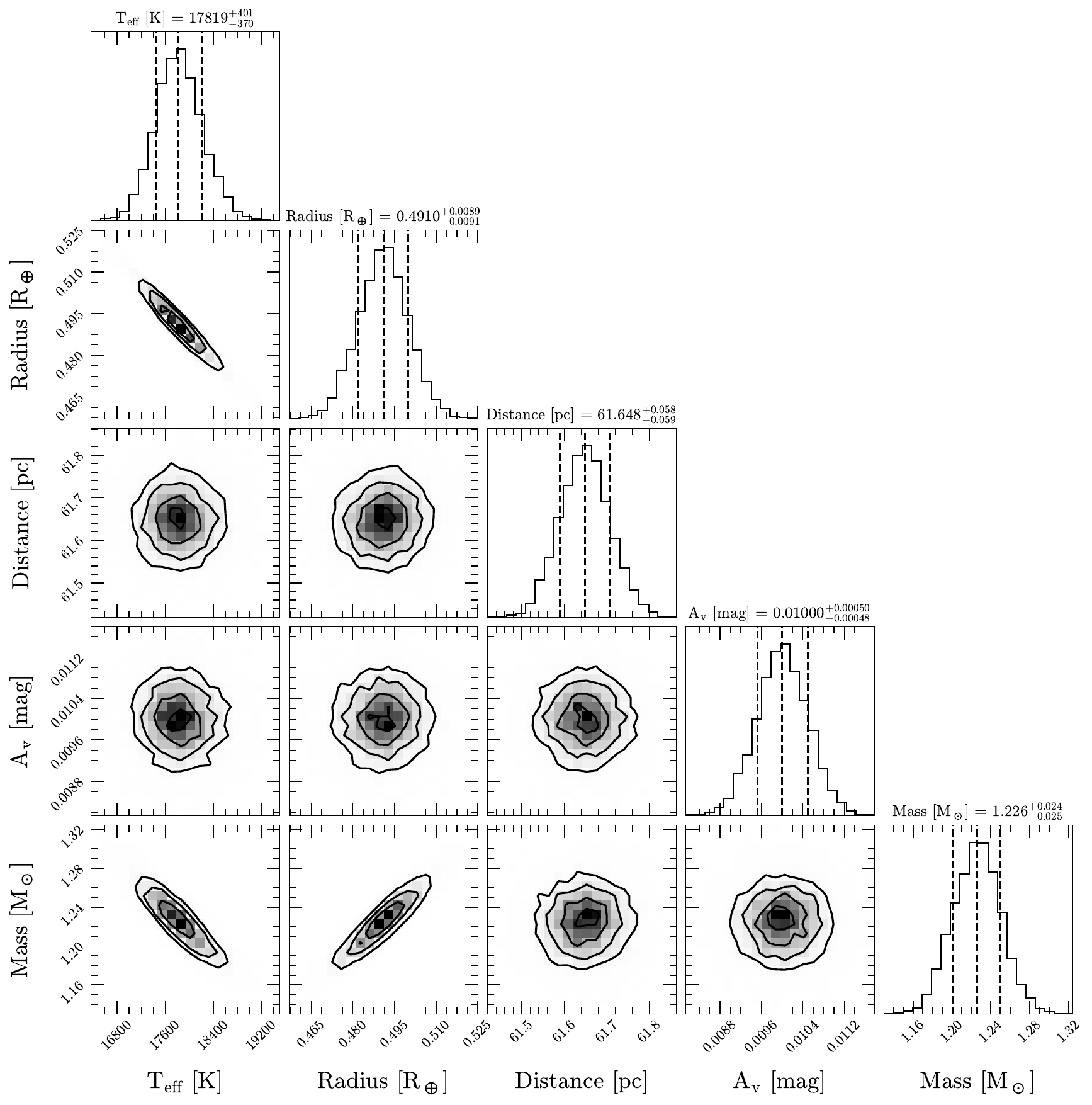}
    \caption{Posterior distributions for the effective temperature, radius, distance, extinction, and mass of SDSS J0608$-$0059 from fits to the SDSS, PanSTARRs, SkyMapper, and Gaia spectral energy distribution.}
    \label{fig:corner}
\end{figure*}



\end{document}